\documentstyle[aps,psfig]{revtex}

\begin{document}

\title{On the Dyadosphere of Black Holes}
\author{Giuliano Preparata}
\address{Physics Department, University and INFN-Section of Milan, Via Celoria 16, 
20133 Milan, Italy}
\author{Remo Ruffini}
\address{I.C.R.A.-International Center for Relativistic Astrophysics and
Physics Department, University of Rome ``La Sapienza", 00185 Rome, Italy}
\author{She-Sheng Xue}
\address{I.C.R.A.-International Center for Relativistic Astrophysics c/o
Physics Department, University of Rome ``La Sapienza", 00185 Rome, Italy}

\maketitle

\begin{abstract}
Basic energy requirements of Gamma Ray Burst(GRB) sources can be easily accounted for by a pair creation process occurring in the ``Dyadosphere" of a Black Hole endowed with an electromagnetic field (abbreviated to EMBH for ``electromagnetic Black Hole"). This includes the recent observations of GRB971214 by Kulkarni {\it et al}. The ``Dyadosphere" is defined as the region outside the horizon of an EMBH where the electromagnetic field exceeds the critical value for $e^+ e^-$ pair production. In a very short time $\sim O({\hbar\over mc^2})$, very large numbers of pairs are created there. Further evolution then leads naturally to a relativistically expanding pair-electromagnetic-pulse (PEM-pulse). Specific examples of Dyadosphere parameters are given for $10M_{\odot}$ and $10^5M_{\odot}$ EMBH's. This process does occur for EMBH with charge-to-mass ratio larger than $2.2\cdot 10^{-5}$ and strictly smaller than one.
From a fundamental point of view, this process represents the first mechanism proved capable of extracting large amounts of energy from a Black Hole with an extremely high efficiency (close to $100\%$).
\end{abstract}
 
\vglue 1cm

It is now generally accepted that Black Holes are uniquely characterized by their total 
mass-energy $E$, charge $Q$, and angular momentum $L$\cite{rw}. The uniqueness theorem has finally been proved after some twenty five years of mathematical work by numerous authors recently reviewed in
\cite{c}. 
The Christodoulou-Ruffini mass formula for Black Holes gives\cite{rc}
\begin{eqnarray}
E^2&=&M^2c^4=\left(M_{\rm ir}c^2 + {Q^2\over2\rho_+}\right)^2+{L^2c^2\over \rho_+^2},\label{em}\\
S&=& 4\pi \rho_+^2=4\pi(r_+^2+{L^2\over c^2M^2})=16\pi\left({G^2\over c^4}\right) M^2_{\rm ir},
\label{sa}
\end{eqnarray}
with
\begin{equation}
{1\over \rho_+^4}\left({G^2\over c^8}\right)\left( Q^4+4L^2c^2\right)\leq 1,
\label{s1}
\end{equation}
where $M_{\rm ir}$ is the irreducible mass, $r_{+}$ is the horizon radius, $\rho_+$ is the quasi spheroidal cylindical coordinate of the horizon evaluated at the equatorial plane,
$S$ is the horizon surface area, and extreme Black Holes satisfy the equality in eq.~(\ref{s1}). [We shall use c.g.s.~units.] From eqs.~(\ref{em}) and (\ref{s1}) it follows that
up to 29$\%$ of the mass-energy of an extreme rotating Black Hole with angular momentum $L_{\rm max}=2\rho^2_{+}c^3/G$ can be stored in its rotational energy and gedanken experiments have been conceived to extract such energy\cite{q,5}. Other processes of rotational energy extraction of astrophysical interest based on magnetohydrodynamic mechanism occurring around a rotating Black Hole have also been advanced\cite{5} though their reversibility as defined in ref.\cite{rc}, and consequently their efficiency of energy extraction is hard to assess precisely.
In the case of Black Holes endowed with an electromagnetic field (EMBH's), it follows from the same equations that up to 50$\%$ of the mass energy of an extreme EMBH with $Q_{\rm max}=\rho_+c^2/\sqrt{G}$ can be stored in its electromagnetic field. We will give here details of a process which can reach almost total reversibility, in the sense of ref.\cite{rc}, and very high efficiency of energy extraction from a EMBH and can be the energy source of GRB's.

By applying the classic work of Heisenberg and Euler \cite{he} as reformulated in a relativistic-invariant form by Schwinger\cite{s},
Damour and Ruffini\cite{dr} showed that a large fraction of the energy of an EMBH can be extracted by pair creation. This energy extraction process can only work for
EMBH black holes with $M_{\rm ir}\lesssim 10^6M_{\odot}$. They also claimed that such an energy source may lead to a natural explanation for GRB's and for ultra high energy cosmic rays.

The recent observations of the Beppo-SAX satellite, the discovery of a very regular afterglow 
to GRB's, the fact their x-ray flux varies regularly with time according to precise power laws\cite{exp}, and especially the optical identification of these sources (see e.g.~ref.\cite{exp}, for GRB971214 ref.\cite{k} and in the infrared by ref.\cite{anr})
have established the correct cosmological setting of  GRB's and have determined their formidable energy requirements. All  these reasons have motivated us to reconsider the above theoretical results and to develop a detailed model for a direct confrontation with the observational results. 

The general considerations presented in Damour and Ruffini\cite{dr} are correct. However,
that work has an underlying assumption which only surfaces in the very last formula: that the pair created in the process of vacuum polarization is absorbed by the EMBH. That view is now fundamentally modified here by the introduction of the novel concept of the Dyadosphere of an EMBH\cite{r}
and by the considerations which follow from its introduction.

For simplicity we use the nonrotating Reissner-Nordstrom EMBH to illustrate the basic gravitational-electrodynamical process. The details on the reversibility, in the sense of Christodoulou-Ruffini\cite{rc}, of pair creation process in the Dyadosphere as well as the breaking of spherical symmetry due to presence of rotation and of an axilly symmetric magnetic field will be presented elsewhere\cite{dr1}.
	
It is appropriate to note that even in the case of an extreme EMBH the charge-to-mass ratio is $10^{18}$ smaller than the typical charge-to-mass ratio found in nuclear matter, owing to the different strengths and ranges of the nuclear and gravitational interactions. This implies that for an EMBH to be extreme, it is enough to have a one quantum of charge present for every $10^{18}$ nucleons in the collapsing matter. 

\begin{figure}[t]
\centerline{
\psfig{figure=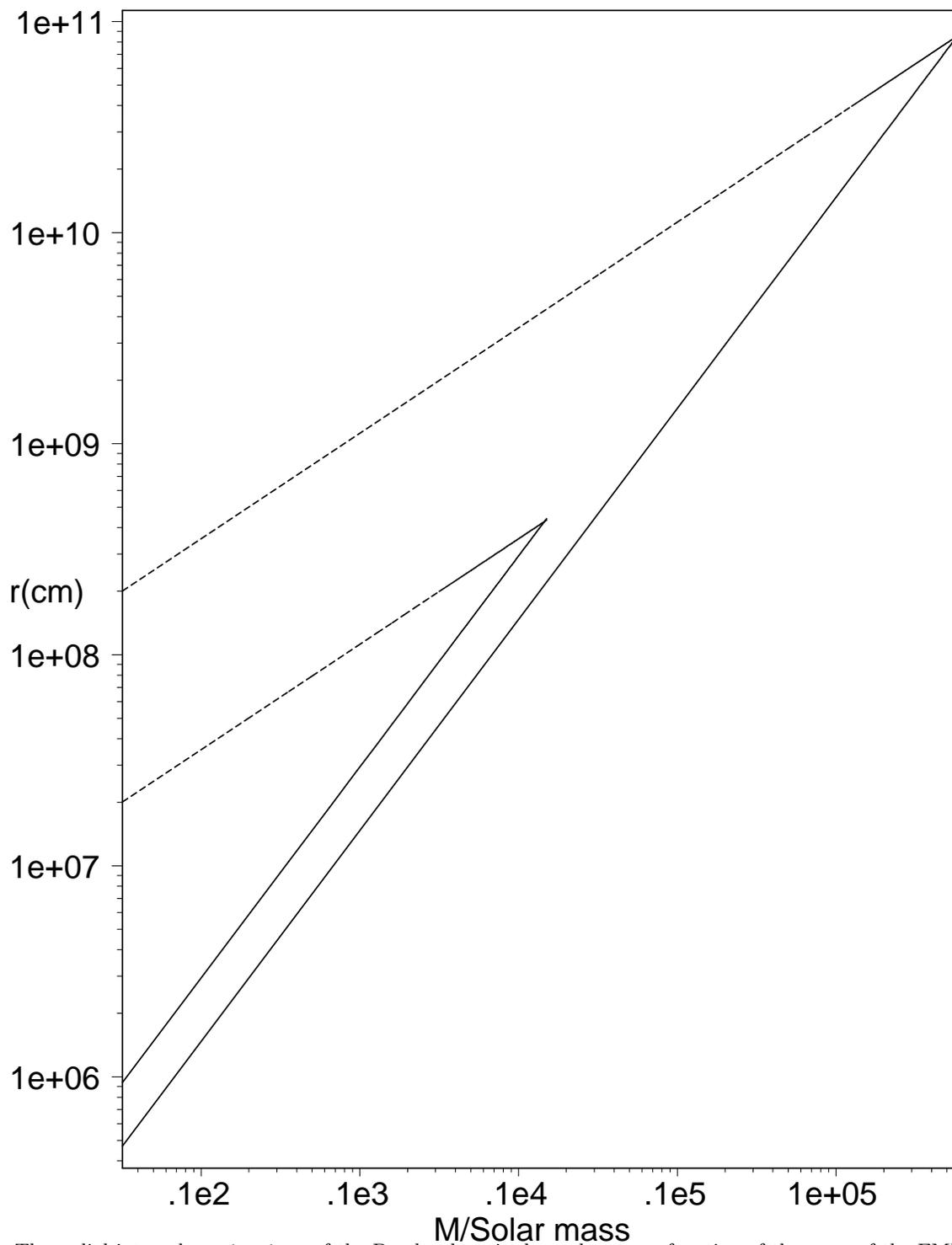}
}
\caption[]{The radial interval $r_+\leq r \leq r_{\rm ds}$ of the Dyadosphere  is shown here as a function of the mass of  the EMBH in solar mass units for the charge parameter values $\xi=1$ (upper curve pair) and $\xi=0.01$ (lower curve pair).
The continuous lines  correspond to the horizon radius $r_+$ given in eq.~(\ref{r+}), and the dotted lines to the Dyadosphere radius $r_{\rm ds}$ given in eq.~(\ref{rc}).}
\label{fig: fig1}
\end{figure}

By introducing the dimensionless mass and charge parameters $\mu={M\over M_{\odot}}>3.2$, $\xi={Q\over Q_{\rm max}}\le 1$, the horizon radius may be expressed as
\begin{equation}
r_{+}={GM\over c^2}\left[1+\sqrt{1-{Q^2\over GM^2}}\right]
=1.47\cdot 10^5\mu (1+\sqrt{1-\xi^2})\hskip0.1cm {\rm cm}.
\label{r+}
\end{equation}
Outside the horizon the electromagnetic field measured in the orthonormal tetrad of an observer at rest at a given point $r$ in the usual Boyer-Lindquist coordinates (see e.g.~\cite{gr}) has only one nonvanishing component ${\bf E}= {Q\over r^2}\hat{r}$ 
along the radial direction. We can evaluate the radius $r_{\rm ds}$ at which the electric field strength
reaches the critical value ${\cal E}_{\rm c}={m^2c^3\over\hbar e}$ introduced by Heisenberg and Euler, where
$m$ and $e$ are the mass and charge of the electron.  
This defines the outer radius of the Dyadosphere, which extends down to the horizon and within which the electric field strength exceeds the critical value. Using the Planck charge $q_{\rm c}= (\hbar c)^{1\over2}$ and the Planck mass $m_{\rm p}=({\hbar c\over G})^{1\over2}$, we can express this outer radius
in the form
\begin{equation}
r_{\rm ds}=\left({\hbar\over mc}\right)^{1\over2}
\left({GM\over c^2}\right)^{1\over2} 
\left({m_{\rm p}\over m}\right)^{1\over2}
\left({e\over q_{\rm p}}\right)^{1\over2}
\left({Q\over\sqrt{G} M}\right)^{1\over2}
=1.12\cdot 10^8\sqrt{\mu\xi} \hskip0.1cm {\rm cm},
\label{rc}
\end{equation} 
which clearly shows the hybrid gravitational and quantum nature of this quantity. The radial interval $r_{+}\leq
r \leq r_{\rm ds}$ describing the Dyadosphere as a function of the  mass is illustrated in Fig.~\ref{fig: fig1} for selected values of the charge paramter $\xi$. 
It is important to note that the Dyadosphere radius is maximized
for the extreme case $\xi=1$ and that the region exists for EMBH's with mass larger than the upper limit for neutron stars, namely $\sim 3.2M_{\odot}$ all the way up to a maximum mass of $6\cdot 10^5M_{\odot}$.
Correspondingly smaller values of the maximum mass are obtained for values of $\xi=0.1,0.01$ as indicated in this figure. For EMBH's with mass larger than the maximum value stated above, the electromagnetic field (whose strength decreases inversely with the mass) never becomes critical. 

We turn now to the crucial issue of the number and energy densities of pairs created in the Dyadosphere. If we consider a shell of proper thickness $\delta\ll {MG\over c^2}$, the electric field is aproximately constant. We can then at each value of the radius $r$ model the electric field as created by a capacitor of width $\delta$ and surface charge density, 
\begin{equation}
\sigma(r)={Q\over 4\pi r^2}.
\label{sq}
\end{equation}
If we now turn to the process of pair creation in order to apply the QED results, we assume $\delta={\hbar\over mc}$ and we can express the rate of pair creation at a given radius $r$ by \cite{he,s}
\begin{equation}
{dN\over\sqrt{-g} d^4x}= {1\over4\pi c}\left({eE\over\pi\hbar}\right)^2
e^{-{\pi E_c\over E}}={1\over4\pi c}\left({4e\sigma\over\hbar}\right)^2
e^{-{\pi \sigma_c\over \sigma}},
\label{rate1}
\end{equation}
where $E=4\pi\sigma$, $\sigma_c={1\over4\pi }{\cal E}_{\rm c}$ is the critical surface charge density and $\sqrt{-g} d^4x$ is 
the invariant four volume. We have for each value of the radius $r$, the rate of pair creation per unit proper time
\begin{equation}
{dN\over d\tau}= {1\over4\pi c}\left({4e\sigma\over\hbar}\right)^2
e^{-{\pi \sigma_c\over \sigma}}4\pi r^2\left({\hbar\over mc}\right).
\label{rate2}
\end{equation}

The pair creation process will continue until a value of the field $E\simeq {\cal E}_{\rm c}$ is reached, or correspondingly, in the capacitor language, till the surface charge density reaches the critical value $\sigma_c$. For $E<{\cal E}_{\rm c}$ or $\sigma<\sigma_c$ the pair creation process is exponentially supressed. We then have
\begin{equation}
\sigma-\sigma_c ={e\over4\pi r^2}{\Delta N\over \Delta \tau}\Delta \tau.
\label{c}
\end{equation}
Based on eq.(\ref{rate2}), one can get,
\begin{equation}
\Delta \tau= {\sigma-\sigma_c\over {e\over4\pi c}\left({4e\sigma\over\hbar}\right)^2
e^{-{\pi \sigma_c\over \sigma}}\left({\hbar\over mc}\right)}\lesssim 
1.99\left({\hbar\over mc^2\alpha}\right)
=1.7610^{-19}{\rm sec.},
\label{dis}
\end{equation}
where $\alpha={e^2\over4\pi\hbar c}$ is the fine structure constant. ( Details are given in ref.\cite{jr}.) The time given by eq.(\ref{dis}) is so short, that the light travel time is smaller or aproximately equal to the width $\delta$. Under these circumstances the correlation between shells can be approximately neglected, thus we can
justify the aproximation of describing the pair creation process shell by shell.

If we now turn to the Dyadosphere its extension goes from the horizon $r_+$ all the way to a radius $r_{ds}$ where ${Q\over4\pi r^2_{ds}}={\cal E}_{\rm c}$. The inner layer of the first shell of width $\delta$, consisting of charges opposite to the one of EMBH's, will be captured by the horizon $r_+$, leading to a new EMBH with charge $Q_c= 4\pi r_+^2{\cal E}_{\rm c}$. The 
outer layer, of oppositely charged particles, will enter the Dyadosphere. The remaining $\sim (r_{ds}-r_+)/{\hbar\over mc}$ shells will also contribute to the plasma constituting the Dyadosphere. It is noteworty to stress the the number of pair created is {\it not}, as nively expected $N_{e^+e^-}={Q-Q_c\over e}$, but the number is tremendously amplified. In the limit 
$r_{ds}\gg {GM\over c^2}$, we have \cite{jr}
\begin{equation}
N_{e^+e^-}\simeq {Q-Q_c\over e}\left[1+{
(r_{ds}-r_+)\over {\hbar\over mc}}\right].
\label{n}
\end{equation}
We can now estimate from eq.(\ref{c}) the number of pairs created at a given radius $r$ in a shell volume of proper thickness $\delta={\hbar\over mc}$
\begin{equation}
\Delta N={4\pi r^2\over e}(\sigma-\sigma_c)={Q\over e}\left[1-\left({r\over r_{ds}}\right)^2\right],
\label{nn}
\end{equation}
and correspondingly the number density of pairs created as a function of the radial coordinate
\begin{equation}
n_{e^+e^-}(r)={\Delta N\over4\pi r^2\left({\hbar\over mc}\right)}
={Q\over e4\pi r^2\left({\hbar\over mc}\right)}\left[1-\left({r\over r_{ds}}\right)^2\right].
\label{nd}
\end{equation}
As shown in \cite{rwx}, these pairs will leave the vicinity of the Black
Hole by creating an enormous PEM-pulse which expands relativistically
out to infinity. The Black Hole will then be left with a charge $Q_{\rm c}$.

Knowing the initial electrostatic energy density and the final electric field value  
$E_{\rm c}(r)={Q_{\rm c}\over r^2}$ holding after the evolution of the Dyadosphere, we compute the energy density of pairs created as a function of the radial coordinate by evaluating 
difference between the initial and final field configurations in the Dyadosphere
\begin{equation}
\rho_{e^+e^-}={1\over8\pi}(E^2(r)-E_c^2)\ .
\label{es}
\end{equation}  
Their total energy is then
\begin{equation}
E^{\rm tot}_{e^+e^-}={1\over2}{Q^2\over r_+}(1-{r_+\over r_{\rm ds}})(1-
\left({r_+\over r_{\rm ds}}\right)^4).
\label{te}
\end{equation}

\begin{figure}[t]
\centerline{
\psfig{figure=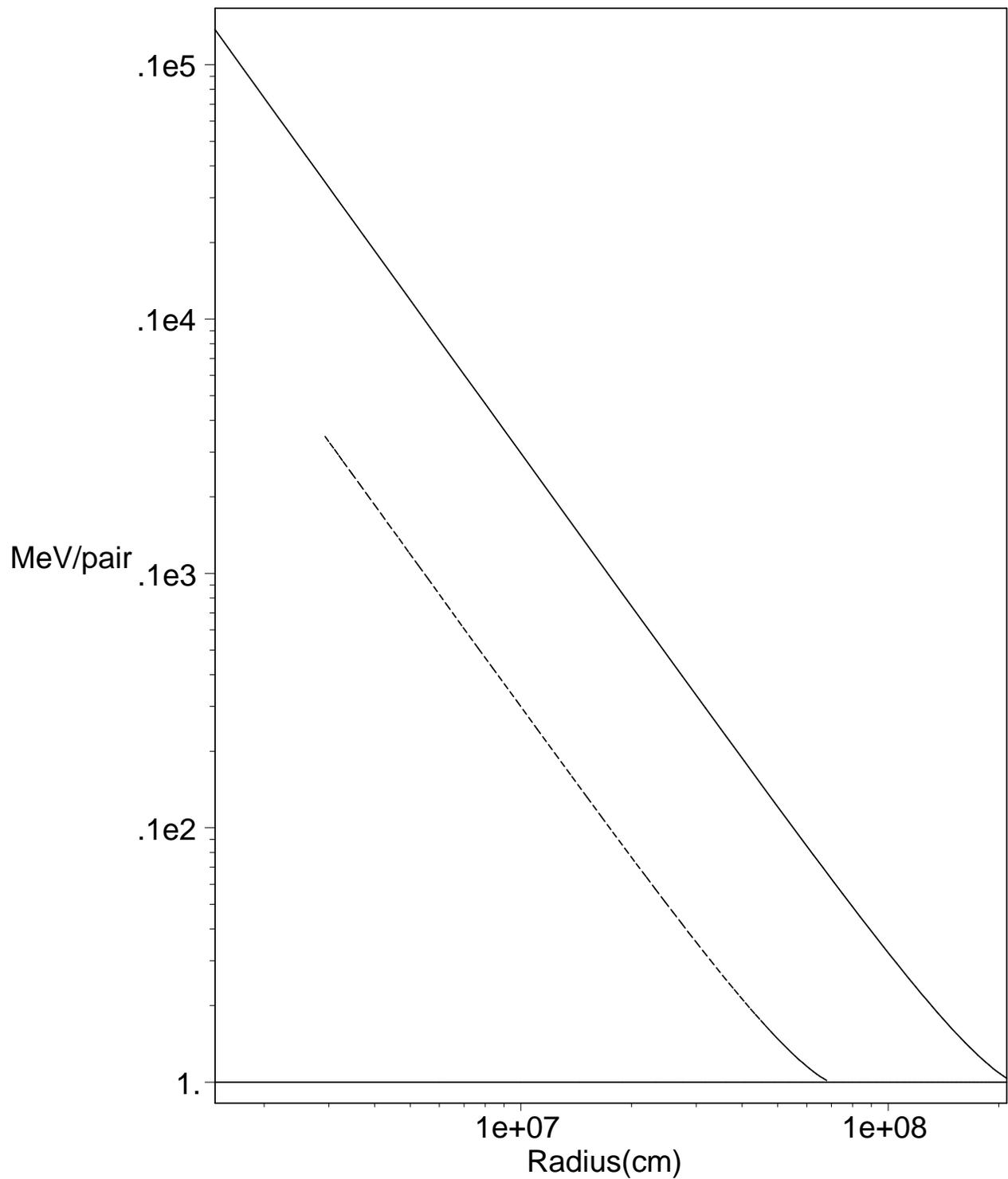}
}
\caption[]{Average energy per pair in MeV plotted as a function of the radius for an EMBH of $10M_\odot$ for the charge parameter values $\xi=1$ (upper curve) and $\xi=0.8$ (lower curve).}
\label{fig: fig2}
\end{figure}

\begin{figure}[t]
\centerline{
\psfig{figure=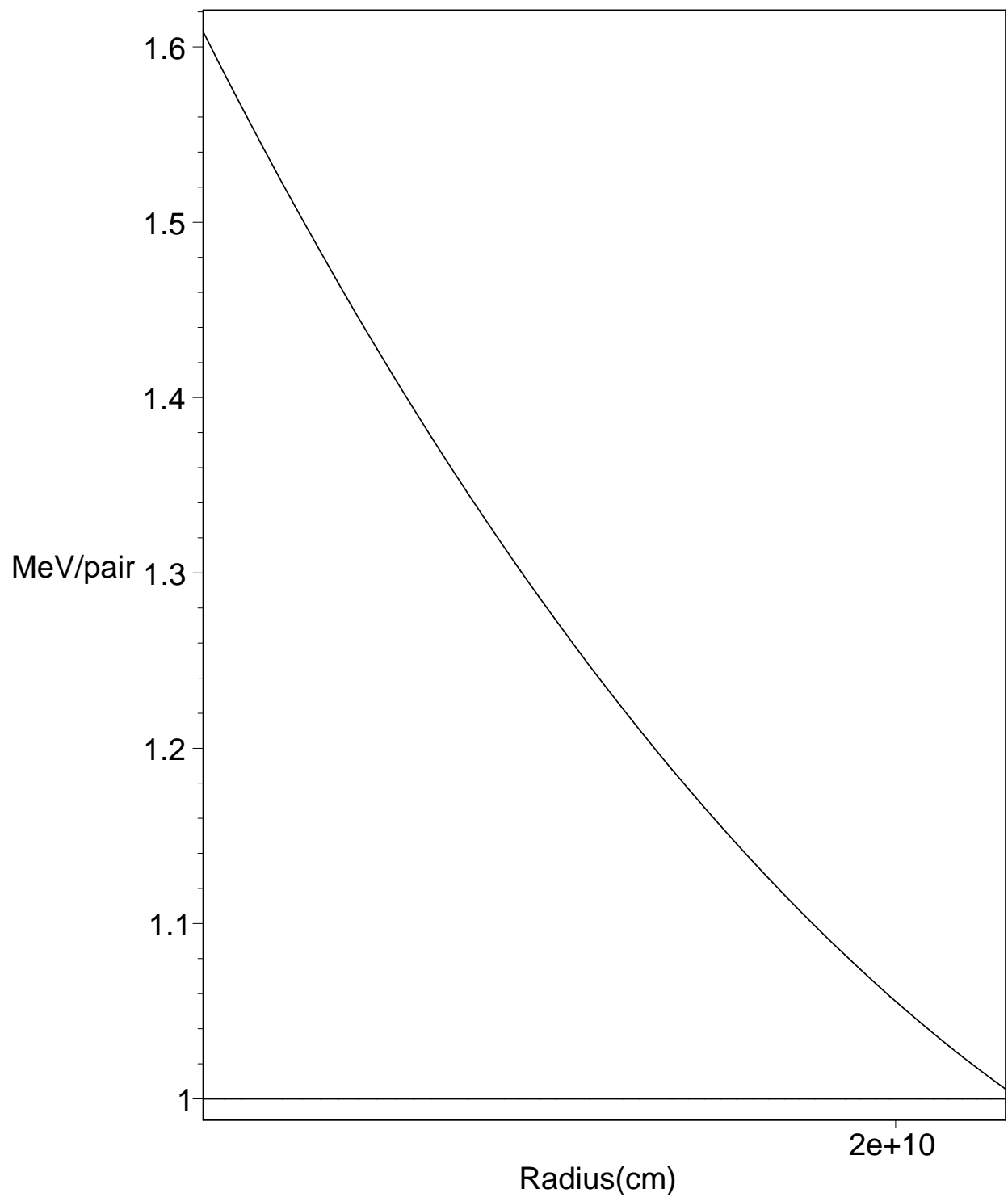}
}
\caption[]{The same as Fig.~\ref{fig: fig2} for an EMBH of $10^5M_\odot$ and $\xi=1$.}
\label{fig: fig3}
\end{figure}

By taking the ratio of eq.(\ref{es}) and eq.(\ref{nd}), we obtain the average energy per pair at each value of the radial coordinate.  Fig.~\ref{fig: fig2} plots this quantity for an
EMBH of $10M_{\odot}$ and Fig.~\ref{fig: fig3} for an EMBH of $10^5M_{\odot}$.
In the first case the energy of pairs near the horizon can reach $10$GeV, while in the second case it never goes over a few MeV. 
These two values of the mass were chosen to be representative of objects typical of the galactic population or for the nuclei of galaxies compatible with our upper limit of the maximum mass of $6\cdot 10^5M_{\odot}$. 

\begin{figure}[t]
\centerline{
\psfig{figure=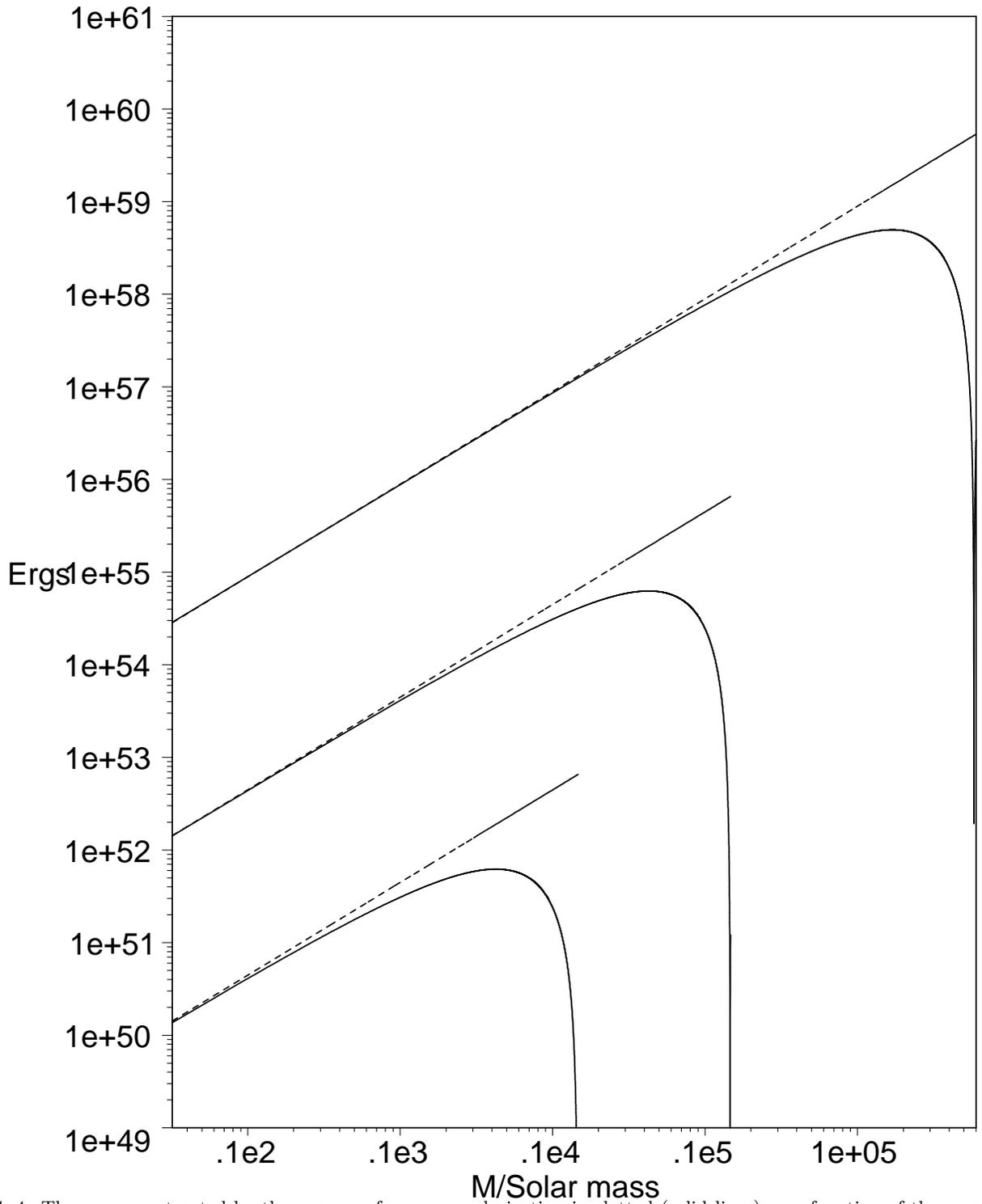}
}
\caption[]{The energy extracted by the process of vacuum polarization is plotted (solid
lines) as a function of the mass $M$ in solar mass units for selected values of the 
charge parameter $\xi=1,0.1,0.01$ for an EMBH, the case $\xi=1$ being reachable only as a limiting process. For comparison we have also plotted the maximum energy extractable from an EMBH (dotted lines) given by eq.~(\ref{em}).}
\label{fig: fig4}
\end{figure}

Finally we can now estimate the total energy extracted by the pair creation process in EMBH's of
different masses for selected values of the charge parameter $\xi$ and compare and contrast these values with the
maximum extractable energy given by the mass formula for Black Holes (see eqs.~(\ref{em}) and 
(\ref{s1})). This comparison is summarized in Fig.~\ref{fig: fig4}. The efficiency of energy extraction by pair creation sharply decreases as the maximum value of the EMBH mass for which vacuum polarization occurs is reached. In the opposite limit the energy of pair creation processes (solid lines in Fig.~\ref{fig: fig4}) asymptotically reaches the entire electromagnetic energy extractable from EMBH's given by eq.(\ref{em}), leading in the limit to fully reversible transformations in the sense of ref.\cite{rc}, $\delta M_{ir}=0$, and $100\%$ efficiency.

It is important to emphasize that this process works for $E(r_+)>{\cal E}_c$. This, in turn, implies that the charge-to-mass ratio for the EMBH ranges between $6.9\cdot 10^{-6}\mu\leq {Q\over\sqrt{G}M}<1$ or $2.2\cdot 10^{-5}\leq {Q\over\sqrt{G}M}<1$, assuming for the minimum mass of the EMBH the $3.2 M_\odot$ limit of Rhoades and Ruffini\cite{rr}. The value $Q=\sqrt{G}M$ is excluded and can only be approached in the limit, since no reversible transformations exist for an extreme EMBH as pointed out in Christodoulou\cite{ch} and Christodoulou and Ruffini\cite{rc}.
  
In order to confront the theoretical model with observations, we must first identify
the gross parameters of the energetics and then refine the analysis. The general
energetics requirements of GRB's, estimated from their energy flux and their distance (see e.g.~ref.\cite{exp} and for GRB971214, refs.\cite{k,anr}) can be directly confronted with integrated results given in Fig.~\ref{fig: fig4}. From the resulting limits on
the mass and the charge parameter $\xi$ of EMBH's, we can proceed to estimate the extent of
the Dyadosphere shown in Fig.~\ref{fig: fig1}, the density of pairs created, and finally the energy distribution in the Dyadosphere shown in Fig.~\ref{fig: fig2} for $10M_{\odot}$ and Fig.~\ref{fig: fig3} for $10^5M_{\odot}$ (see also ref.\cite{prx}). 
 
Our model of GRB's, based on the results presented in this letter, is very different from the ones debated in the last ten years in the scientific literature, it can overcome some of their basic difficulties and on some specific aspects it can also have important analogies:

\begin{itemize}

\item	
the EMBH drastically differs from the most popular binary neutron stars model\cite{17} debated in the last 16 years. It presents over those models the distinct advantage of not having the recognized difficulty of explaining the energetic of GRB's\cite{18}, and especially offers the possibility of computing explicitly and uniquely the details of the initial conditions by the energetic of the Dyadosphere as given in the above Fig.~\ref{fig: fig2} and Fig.~\ref{fig: fig3} .

\item
the further evolution of the Dyadosphere (the PEM pulse), has been analyzed  by a variety of theoretical models based on relativistic hydrodynamics equations both on semi- analytical-numerical idealized models as well as with the help of a fully general relativistic hydrodynamical code. The combined results are at variance with many of the considerations in the current published papers dealing with fireballs and relativistic winds\cite{19}, though for a few aspects they present striking analogies. Details are given in ref.\cite{rwx}.

\item 
our above two treatments lead uniquely to the computation of the further evolution of the PEM Pulse in the afterglow era and again these specific analysis which can be uniquely derived from the above assumptions are at variance with the ones in the existing literature\cite{21} though they present some interesting analogies and they appear to be consistent with the latest observations\cite{rwx,22}. 

\end{itemize}

If our basic scenario is confirmed by observation, additional fundamental problems should be addressed: (i) the origin of the Dyadosphere, 
and (ii) the astrophysics of the remnant EMBH's. 

The understanding of the origin of the Dyadosphere implies the solution of two fundamental problems: 
(a) the development of relativistic magneto-hydrodynamical processes
occurring in the accreting material with special attention to processes of charge separation\cite{jw,rt} as well as to the identification of physical processes leading to the formation of a charge depleted region with an electric field sufficient to polarize the vacuum, and
(b) finding the astrophysical conditions leading to
the gravitational	collapse of masses of $M\sim O(10^5M_{\odot})$, if the observations
support the existence of such objects.

The remnant EMBH's that result from such catastrophic events still have such a large amount of stored energy and such electrodynamical properties that
they can be the source of the ultra high energy tail of cosmic rays ($E\geq 10^{16}$eV) as originally suggested\cite{dr,dr1}. An entire new problematic arises if the remnant is accreting, leading to the interesting possibility for a new mechanism for the formation of jets.

{\it Note added}: This paper presented a fundamental concept of the dyadosphere
of black holes. This work and ref.\cite{prx} provide a fundamental basis for the theoretical model\cite{rwx,ruffinietal} explaning Gamma-Ray-Bursts phenomenon.

\newpage 

%

\newpage 

\newpage 

\newpage 

\end{document}